\documentclass[journal=jctcce,manuscript=article]{achemso}
\usepackage{graphicx,bm,microtype,hyperref,multirow,amscd,ifsym,xcolor,braket,footnote}
\usepackage[version=4]{mhchem}
\newcommand{\alert}[1]{\textcolor{black}{#1}}
\newcommand{\mc}{\multicolumn}

\DeclareMathOperator{\erfc}{erfc}
\newcommand{\purple}[1]{\textcolor{purple}{#1}}
\newcommand{\orange}[1]{\textcolor{orange}{#1}}
\newcommand{\br}{\mathbf{r}}
\newcommand{\bo}{\mathbf{0}}
\newcommand{\ba}{\mathbf{a}}
\newcommand{\bb}{\mathbf{b}}
\newcommand{\bs}{\mathbf{s}}
\newcommand{\bA}{\mathbf{A}}
\newcommand{\bB}{\mathbf{B}}
\newcommand{\bY}{\mathbf{Y}}
\newcommand{\bZ}{\mathbf{Z}}
\newcommand{\BM}{\mathbf{m}}
\newcommand{\lam}{\lambda}

\author{Giuseppe M. J. Barca}
\author{Pierre-Fran\c{c}ois Loos}
\email{pf.loos@anu.edu.au}
\author{Peter M. W. Gill}
\email{peter.gill@anu.edu.au}
\affiliation{Research School of Chemistry, Australian National University, ACT 2601, Australia}

\title{Many-electron integrals over gaussian basis functions. \\
	I. Recurrence relations for three-electron integrals}
\begin{document}

\begin{abstract}
Explicitly-correlated F12 methods are becoming the first choice for high-accuracy molecular orbital calculations, and can often achieve chemical accuracy with relatively small gaussian basis sets. 
In most calculations, the many three- and four-electron integrals that formally appear in the theory are avoided through judicious use of resolutions of the identity (RI).  However, in order not to jeopardize the intrinsic accuracy of the F12 wave function, the associated RI auxiliary basis set must be large.
Here, inspired by the Head-Gordon-Pople (HGP) and PRISM algorithms for two-electron integrals, we present an algorithm to compute directly three-electron integrals over gaussian basis functions 
and a very general class of three-electron operators, without invoking RI approximations.
A general methodology to derive vertical, transfer and horizontal recurrence relations is also presented. 
\end{abstract}

\section{Introduction}
Many years ago, Kutzelnigg \cite{Kutzelnigg85} showed that introducing the interelectronic distance $r_{12} = |\br_1- \br_2|$ into a simple wave function $\Psi_0$ for the helium atom, to yield
\begin{equation}
	\Psi = \left(1 + \frac{r_{12}}{2} \right) \Psi_0 + \chi(\br_1,\br_2),
\end{equation}
where $\chi(\br_1,\br_2)$ is expanded in a determinantal basis, dramatically improves the convergence of second-order perturbation theory, \cite{Kutzelnigg91} coupled cluster \cite{Noga94} and other variational calculations. \cite{Klopper91b}
This approach, which gave birth to the so-called ``R12 methods'', \cite{Kutzelnigg91, Termath91, Klopper91a} is a natural extension of pioneering work by Hylleraas in the 1920s. \cite{Hylleraas28, Hylleraas29}

Kutzelnigg's idea was later generalized to more accurate correlation factors $f_{12} \equiv f(r_{12})$. \cite{Persson96, Persson97, May04, Tenno04b, Tew05, May05}
Nowadays, a popular choice for $f_{12}$ is a Slater-type geminal \cite{Tenno04a} (STG)
\begin{equation}
	f_{12}^\text{STG} = \exp(-\lam\,r_{12}),
\end{equation}
which is sometimes expanded as a linear combination of gaussian-type geminals \cite{Persson96, Persson97} (GTG)
\begin{equation}
	f_{12}^\text{STG} \approx \sum_{k=1}^{N_\text{G}} c_k\,f_{12}^\text{GTG}(\lam_k) = \sum_{k=1}^{N_\text{G}} c_k\exp(-\lam_k\,r_{12}^2) 
\end{equation}
for computational convenience. \cite{May04, May05, Tew05} 

The resulting ``F12 methods'' achieve chemical accuracy for small organic molecules with relatively small gaussian basis sets \cite{Klopper06, Hattig12, Kong12,Tenno12a, Tenno12b} and are quickly becoming the first-choice non-stochastic method for high accuracy.\cite{Tenno12a, Tenno12b} They are now appearing within state-of-the-art composite procedures. \cite{Karton12}

However, whereas a conventional (non-explicitly correlated) calculation using $N$ gaussian one-electron basis functions leads to an energy expression with $O(N^4)$ relatively easy two-electron integrals, \cite{Boys50, Pople78, McMurchie78, Rys83, Obara86, Obara88, HGP88, Braket90, ssss91, PRISM91, Review94} the explicitly correlated wave function has an energy expression with $O(N^6)$ three-electron integrals and $O(N^8)$ four-electron integrals.
These many-electron integrals are known in closed form in a few special cases \cite{Wind01} but, in general, they are computationally expensive and the task of computing $O(N^6)$ or $O(N^8)$ of them appears overwhelming.

To render F12 methods computationally tractable and to avoid these three- and four-electron integrals without introducing unacceptable errors, Kutzelnigg and Klopper \cite{Kutzelnigg91} proposed to insert the {\em resolution of the identity} (RI)
\begin{equation} \label{eq:RI}
	\hat{I} \approx \sum_{\mu}^{N_\text{RI}} \left| \chi_\mu \right> \left<  \chi_\mu \right|
\end{equation}
into the three- or four-electron operators. \cite{Werner07, Hattig12}
In this way, three- and four-electron integrals are expanded in terms of one- and two-electron integrals that can be found efficiently using conventional quantum chemistry software.

Of course, the approximation \eqref{eq:RI} introduces chemically acceptable errors only if the auxiliary basis set $\{ \chi_{\mu} \}$ is sufficiently large ($N_\text{RI} \gg N$).
 \footnote{For example, for the leflunomide molecule \ce{C12H9F3N2O2} studied recently by Bachorz et al.\cite{Bachorz11}, the primary aug-cc-pVTZ and auxiliary  \cite{Yousaf08, Yousaf09} basis sets have 1081 and 2864 basis functions, respectively.}
This realization has fueled much research and there now exist very efficient RI formulations based on highly optimized auxiliary basis sets (ABS) \cite{Klopper02} and complementary ABS (CABS). \cite{Valeev04}  However, to achieve millihartree accuracy, such bases must be approximately complete up to $3L_{occ}$,\cite{Hattig12} where $L_{occ}$ is the highest angular momentum of basis functions in the orbital basis set associated with the occupied molecular orbitals.  Even if this requirement can be reduced to $2L_{occ}$ \alert{by density fitting,\cite{Manby2003}} it remains demanding in cases (e.g. for transition metals) where high angular momenta are needed in the orbital basis.

For two reasons, however, we believe that the RI tactic can be avoided.  First, the number of \emph{significant} (i.e.~greater than a threshold) three- and four-electron integrals is very much smaller than $O(N^6)$ and $O(N^8)$, respectively.  Second, the task of computing many-electron integrals over gaussian basis functions is less formidable than many believe. \alert{For example, for three-electron integrals, a reduction of the computational effort from $O(N^6)$ to $O(N^4)$ is already achievable by exploiting robust density fitting techniques.\cite{Manby2014}}

Although there are $O(N^4)$ two-electron integrals, it is well known to quantum chemistry programmers that the number of \emph{significant} two-electron integrals in a large system is only $O(N^2)$ if the two-electron operator is long-range \cite{Ahlrichs74, Haser89, Bound94} and $O(N)$ if it is short-range.\cite{CAP96, Eshort99}
Similar considerations apply to many-electron integrals and one can show that the number of \emph{significant} three- and four-electron integrals is only $O(N^3)$ and $O(N^4)$, respectively, even for long-range operators.  (For short-range operators, there are even fewer.)  Thus, if we can find good algorithms for identifying and computing the tiny fraction of many-electron integrals that are significant, large-scale calculations using F12 methods will become feasible, without the need for the RI approximation.

The present manuscript is the first of a series on many-electron integrals.
Here, we present an algorithm to compute integrals over three-electron operators using recurrence relations (RRs). The second paper \cite{3ERI2} in the series will discuss the construction of upper bounds and effective screening strategies and the third \cite{3ERI3} will describe optimized numerical methods for the efficient generation of $[sss|sss]$ integrals. 
Our recursive approach applies to a general class of multiplicative three-electron operators and thus generalizes existing schemes that pertain only to GTGs. \cite{Persson97, Tenno00, Saito01,DahleThesis, Dahle2007,Dahle2008, Komornicki11}

Section \ref{sec:integrals} contains basic definitions, classifications of three-electron operators, and permutational symmetry considerations.
In Sec.~\ref{sec:algorithm}, we propose a recursive algorithm for the computation of three-electron integrals. 
Details of a general scheme for deriving three-electron integral RRs are presented in the Appendix.
Atomic units are used throughout.

\section{\label{sec:integrals} Three-electron integrals}
A primitive gaussian-type orbital (PGTO)
\begin{equation}
	\varphi_{\ba}^{\bA}(\br) = (x-A_x)^{a_x} (y-A_y)^{a_y} (z-A_z)^{a_z} e^{-\alpha \left| \br-\bA \right|^2}
\end{equation}
is defined by its exponent $\alpha$, its center $\bA=(A_x,A_y,A_z)$, its angular momentum vector $\ba = (a_x,a_y,a_z)$ and its total angular momentum $a = a_x+a_y+a_z$.

A contracted gaussian-type orbital (CGTO) $\psi_{\ba}^{\bA}(\br)$ is a linear combination of $K_A$ PGTOs.

We are interested in the three-electron operator 
\begin{equation}
	f_{12}\,g_{13}\,h_{23} \equiv f(r_{12})\,g(r_{13})\,h(r_{23}),
\end{equation}
and we write the integral of six CGTOs over this as
\begin{align}
	\braket{\ba_1\ba_2\ba_3|\bb_1\bb_2\bb_3}	& \equiv \braket{\ba_1\ba_2\ba_3 | f_{12} g_{13} h_{23} | \bb_1\bb_2\bb_3}							\notag	\\
												& = \iiint \psi_{\ba_1}^{\bA_1}(\br_1) \psi_{\ba_2}^{\bA_2}(\br_2) \psi_{\ba_3}^{\bA_3}(\br_3) \,f_{12} \,g_{13} \,h_{23} \,
														\psi_{\bb_1}^{\bB_1}(\br_1) \psi_{\bb_2}^{\bB_2}(\br_2) \psi_{\bb_3}^{\bB_3}(\br_3)	d \br_1 d \br_2 d \br_3.
\end{align}
We will use square brackets if the integral is over PGTOs:
\begin{align}
	[\ba_1\ba_2\ba_3|\bb_1\bb_2\bb_3]		& \equiv [\ba_1\ba_2\ba_3 | f_{12}g_{13}h_{23}  | \bb_1\bb_2 \bb_3]								\notag	\\
												& = \iiint \varphi_{\ba_1}^{\bA_1}(\br_1) \varphi_{\ba_2}^{\bA_2}(\br_2) \varphi_{\ba_3}^{\bA_3}(\br_3) \,f_{12}\,g_{13}\,h_{23} \,
														\varphi_{\bb_1}^{\bB_1}(\br_1) \varphi_{\bb_2}^{\bB_2}(\br_2) \varphi_{\bb_3}^{\bB_3}(\br_3) d \br_1 d \br_2 d \br_3,
\end{align}
and the fundamental integral (i.e.~one in which all six PGTOs are $s$-type) is therefore
\begin{equation} \label{eq:def3}
	[\bo\bo\bo|\bo\bo\bo] = S_1 S_2 S_3 \iiint \varphi_{\bo}^{\bZ_1}(\br_1) \varphi_{\bo}^{\bZ_2}(\br_2) \varphi_{\bo}^{\bZ_3}(\br_3) \,f_{12}\,g_{13}\,h_{23} \,d\br_1 d\br_2 d\br_3,
\end{equation}
where the exponents $\zeta_i$, centers $\bZ_i$ and prefactors $S_i$ of the gaussian product rule for the three electrons are
\begin{gather}
	\zeta_i = \alpha_i + \beta_i,										\\
	\bZ_i = \frac{\alpha_i \bA_i + \beta_i \bB_i}{\alpha_i + \beta_i},	\\
	S_i = \exp\left[-\frac{\alpha_i \beta_i}{\alpha_i+\beta_i} |\bA_i-\bB_i|^2 \right].
\end{gather}
For conciseness, we will adopt a notation in which missing indices represent s-type gaussians.  For example, $[\ba_2\ba_3]$ is a shorthand for $[\bo\ba_2\ba_3 | \bo\bo\bo]$.  We will also use unbold indices, e.g. $\braket{a_1a_2a_3|b_1b_2b_3}$ to indicate a complete class of integrals from a shell-sextet.

\subsection{Three-electron operators}
We are particularly interested in two types of three-electron operators:
``chain'' operators of the form $f_{12}\, g_{13}$, and ``cyclic'' operators of the form $f_{12}\,g_{13}\,h_{23}$.
In both types, the most interesting cases arise when \cite{Klopper06, Kutzelnigg91, Hattig12, Kong12}
\begin{equation*}
	f_{12},\,g_{12},\,h_{12} =	\begin{cases}
									r_{12}^{-1},					&	\text{Coulomb operator},		\\
									r_{12},						&	\text{anti-Coulomb operator},	\\
									\exp(-\lam\,r_{12}),			&	\text{Slater-type geminal},		\\
									\exp(-\lam\,r_{12}^{2} ),		&	\text{gaussian-type geminal},
								\end{cases}	
\end{equation*}
and various combinations of these produce three-electron integrals of practical importance.  We note that, by virtue of the identity $r_{12} \equiv r_{12}^{-1} (r_1^2 + r_2^2 - 2\br_1\cdot\br_2)$, integrals involving the anti-Coulomb operator can be reduced to linear combinations of integrals over the Coulomb operator.

\subsection{Permutational symmetry}
For real basis functions, it is known \cite{SzaboBook} that two-electron integrals have 8-fold permutational symmetry, meaning that the integrals
\begin{align*}
	& \braket{\ba_1 \ba_2 | \bb_1 \bb_2}	&& \braket{\bb_1 \ba_2 | \ba_1 \bb_2}		&& \braket{\bb_1 \bb_2 | \ba_1 \ba_2}		&& \braket{\ba_1 \bb_2 | \bb_1 \ba_2}	\\
	& \braket{\ba_2 \ba_1 | \bb_2 \bb_1}	&& \braket{\bb_2 \ba_1 | \ba_2 \bb_1}		&& \braket{\bb_2 \bb_1 | \ba_2 \ba_1}		&& \braket{\ba_2 \bb_1 | \bb_2 \ba_1}
\end{align*}
are all equal.  Three-electron integrals also exhibit permutational symmetry and, for computational efficiency, it is important that this be fully exploited.  The degeneracy depends on the nature of the three-electron operator and the five possible cases are listed in Table \ref{tab:perms}.

\begin{table}	
	\caption{Permutational degeneracy for various operators. \label{tab:perms}}
		\begin{tabular}{lcccc}
			\hline
			\hline
			Type					&	Operator					&	Degeneracy	\\
			\hline
			Two-electron			&	$f_{12}$					&		8			\\[2mm]
			Three-electron chain	&	$f_{12} g_{13}$				&		8			\\
									&	$f_{12} f_{13}$				&		16			\\[2mm]
			Three-electron cyclic	&	$f_{12} g_{13} h_{23}$		&		8			\\
									&	$f_{12} f_{13} h_{23}$		&		16			\\
									&	$f_{12} f_{13} f_{23}$		&		48			\\
			\hline
			\hline
		\end{tabular}	
\end{table}

\section{\label{sec:algorithm} Algorithm}
\begin{figure}
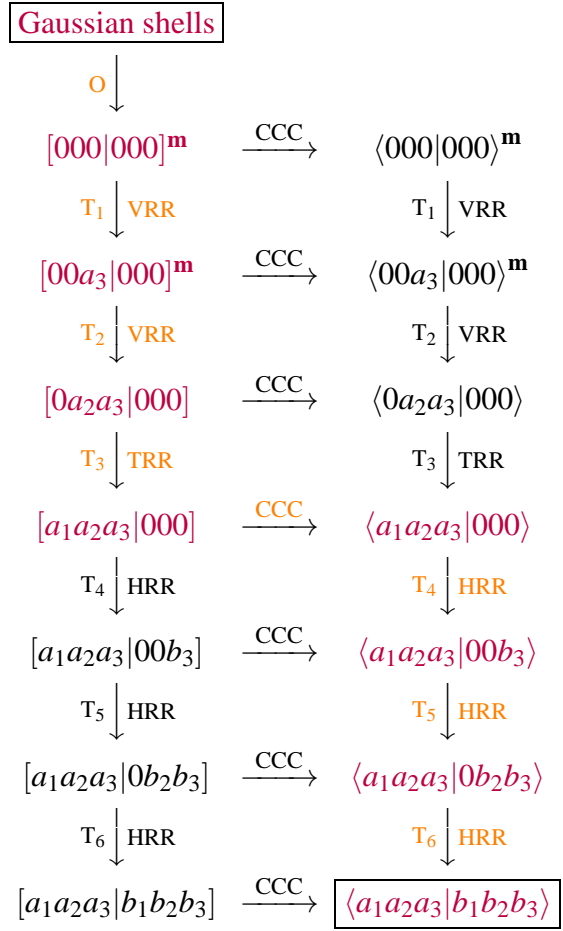
	
$$
\begin{CD}
@.	\boxed{\text{\purple{Gaussian shells}}}																									\\
@.		@V{\orange{\text{O}}}V\orange{}V																									\\
@.			\purple{[000|000]^{\BM}} 						@>\text{CCC}>>					\braket{000|000}^{\BM} 					\\
@.		@V{\orange{\text{T}_1}}V\orange{\text{VRR}}V										@V{\text{T}_1}V\text{VRR}V					\\
@.			\purple{[00a_3|000]^{\BM}}						@>\text{CCC}>>					\braket{00a_3|000}^{\BM}					\\	
@.		@V{\orange{\text{T}_2}}V\orange{\text{VRR}}V										@V{\text{T}_2}V\text{VRR}V					\\
@.			\purple{[0a_2a_3|000]} 							@>\text{CCC}>>					\braket{0a_2a_3|000}						\\
@.		@V{\orange{\text{T}_3}}V\orange{\text{TRR}}V										@V{\text{T}_3}V\text{TRR}V					\\
@.			\purple{[a_1a_2a_3|000]} 					@>\text{\orange{CCC}}>>			\purple{\braket{a_1a_2a_3|000}} 				\\
@.		@V{\text{T}_4}V\text{HRR}V													@V{\orange{\text{T}_4}}V\orange{\text{HRR}}V		\\
@.				[a_1a_2a_3|00b_3]							@>\text{CCC}>>				\purple{\braket{a_1a_2a_3|00b_3}}				\\
@.		@V{\text{T}_5}V\text{HRR}V													@V{\orange{\text{T}_5}}V\orange{\text{HRR}}V		\\
@.				[a_1a_2a_3|0b_2b_3]						@>\text{CCC}>>				\purple{\braket{a_1a_2a_3|0b_2b_3}} 			\\
@.		@V{\text{T}_6}V\text{HRR}V													@V{\orange{\text{T}_6}}V\orange{\text{HRR}}V		\\
@.				[a_1a_2a_3|b_1b_2b_3]					@>\text{CCC}>>		\boxed{\purple{\braket{a_1a_2a_3|b_1b_2b_3}}}		\\
\end{CD}
$$
\caption{
\label{fig:algo}
PRISM representation\cite{PRISM91} of a scheme for computing a three-electron integral class.  In this work, we consider the (orange) OTTTCCCTTT path.}
\end{figure}

In this Section, we present a recursive algorithm for generating a class of three-electron integrals of arbitrary angular momentum from an initial set of fundamental integrals. 
The algorithm applies to any three-electron operator of the form $f_{12}\,g_{13}\,h_{23}$ and generalizes the HGP-PRISM algorithm following a OTTTCCCTTT pathway. \cite{HGP88, PRISM91}  The algorithm is shown schematically in Fig. \ref{fig:algo}. 

After selecting a significant shell-sextet, we create a set of generalized fundamental integrals $[\bo]^{\BM}$ (Step O).
Next, we build angular momentum on center $\bA_3$ (Step T$_1$) and on center $\bA_2$ (Step T$_2$) using vertical RRs (VRRs). 
This choice is motivated by the fact that, for chain operators, the VRR for building momentum on $\bA_1$ is more expensive than that for building on $\bA_2$ and $\bA_3$ (see Appendix).
Then, using transfer RRs (TRRs), we transfer momentum onto $\bA_1$ (Step T$_3$).
The primitive $[\ba_1\ba_2\ba_3]$ integrals are then contracted (Step $\text{C}$) and horizontal RRs (HRRs) are used to shift angular momentum from the bra centers $\bA_3$, $\bA_2$ and $\bA_1$ onto the ket centers $\bB_3$, $\bB_2$ and $\bB_1$ (Steps T$_4$, T$_5$ and T$_6$).  The number of terms in each of these RRs is summarized in Table \ref{tab:RRterm} for cyclic and chain operators.  We now describe each step in detail.

\begin{table}	
	\caption{Number of RR terms for cyclic and chain operators. \label{tab:RRterm}}
		\begin{tabular}{lcccc}
			\hline
			\hline
			Step		&	RR type	&		Expression				&				\mc{2}{c}{Operators}					\\
			\cline{4-5}
						&				&								&	$f_{12}\,g_{13}\,h_{23}$	&	$f_{12}\,g_{13}$	\\
			\hline
			T$_1$		&	VRR		&	Eq.~\eqref{eq:T1_cyclic}	&		8				&				6				\\
			T$_2$		&	VRR		&	Eq.~\eqref{eq:T2_cyclic}	&		10				&				7				\\
			T$_3'$		&	VRR		&	Eq.~\eqref{eq:T3_prime}	&		12				&				12				\\
			T$_3$		&	TRR		&	Eq.~\eqref{eq:T3_cyclic}	&		6				&				6				\\
			T$_4$		&	HRR		&	Eq.~\eqref{eq:T4}			&		2				&				2				\\
			T$_5$		&	HRR		&	Eq.~\eqref{eq:T5}			&		2				&				2				\\
			T$_6$		&	HRR		&	Eq.~\eqref{eq:T6}			&		2				&				2				\\
			\hline
			\hline
		\end{tabular}	
\end{table}

\subsection{Construct shell-pairs, -quartets and -sextets}
	Beginning with a list of shells, a list of significant shell-pairs\cite{Review94} is constructed.  By pairing these shell-pairs, a list of significant shell-quartets is created and then, by pairing the significant pairs and quartets, a list of significant shell-sextets is created.  This process is critical for the efficiency of the overall algorithm and depends on the use of tight upper bounds for the target integrals. Such bounds are straightforward for two-electron integrals\cite{Ahlrichs74, Haser89, Bound94, CAP96, Eshort99} but are much more complicated for three-electron integrals and, for example, the popular Cauchy-Schwartz bound does not generalize easily.  However, it is possible to construct a bound for each type of three-electron operator, depending on the short- or long-range  character of $f_{12}$, $g_{13}$ and (if it is present) $h_{23}$.  We will discuss these in detail in Part II of this series.\cite{3ERI2}

\vspace{3cm}

\begin{table}	
\caption{
\label{tab:kern}
\alert{Laplace kernels $F(s)$ for various two-electron operators $f(r_{12})$.  $n$ is an integer.  $\Gamma$ is the gamma function, $H_n$ is a Hermite polynomial, $\erfc$ is the complementary error function, and $(a)_j$ is a Pochhammer symbol.  $\delta^{(k)}$ and $\theta^{(k)}$ are the kth derivatives of the Dirac delta function and Heaviside step function, respectively.\cite{NISTbook}}}
\begin{tabular}{ccc}
	\hline \hline																																						\\
	$f(r_{12})$										&												$F(s)$															\\	\noalign{\bigskip}
	\hline																																								\noalign{\bigskip}
	$r_{12}^n \exp(-\lam r_{12})$	&	$\displaystyle{\frac{2\pi^{-1/2}}{(4s)^{n/2+1}} H_{n+1}\left(\frac{\lam}{2s^{1/2}}\right) \exp\left(-\frac{\lam^2}{4s}\right)}$		\\	\noalign{\bigskip}
	$(r_{12}^2-R^2)^n \exp(-\lam^2 r_{12}^2)$		&	$\displaystyle{\exp[-R^2(s-\lam^2)] \delta^{(n)}(s-\lam^2)}$													\\	\noalign{\bigskip}
	$(r_{12}^2-R^2)^{n-1/2} \exp(-\lam^2 r_{12}^2)$&	$\displaystyle{\frac{\exp[-R^2(s-\lam^2)] \theta(s-\lam^2)}{\Gamma(-n+1/2)(s-\lam^2)^{n+1/2}}}$			\\	\noalign{\bigskip}
	$r_{12}^{2n} \erfc(\lam r_{12})$					&	$\displaystyle{\frac{-\pi^{-1/2}}{\Gamma(-n-1/2) s^{n+1}}
															\sum_{k=0}^n \binom{n}{k} \left(\frac{\lam^2}{s-\lam^2}\right)^{k+1/2} \frac{\theta(s-\lam^2)}{k+1/2}}$	\\	\noalign{\bigskip}
	$r_{12}^{2n-1} \erfc(\lam r_{12})$				&	$\displaystyle{\sum_{k=0}^n \binom{n}{k} \frac{s^{k-n-1/2}}{\Gamma(k-n+1/2)} \theta^{(k)}(s-\lam^2)}$		\\	\noalign{\bigskip}
	$r_{12}^{2n} \erfc(\lam r_{12})^{2}$				&	$ \displaystyle{\frac{2\lam}{\pi s\sqrt{s-\lam^2}} \sum_{k=0}^n \binom{n}{k} (3/2)_k
															\frac{\theta^{(n-k)}(s-2\lam^2)}{(\lam^2-s)^k} \sum_{j=0}^k \frac{(-k)_j}{(3/2)_j} (\lam^2/s)^j}$	\\	\noalign{\bigskip}
	\hline \hline
\end{tabular}	
\end{table}

\subsection{Step O.  Form fundamental integrals}
Having chosen a significant shell-sextet, we replace the three two-electron operators in its fundamental integral \eqref{eq:def3} by their Laplace representations
\begin{subequations}
\begin{gather}
	f(r_{12}) = \int_0^\infty F(s_1)  \exp{\left(-s_1 r_{12}^2\right)} \,ds_1,	\\ 
	g(r_{13}) = \int_0^\infty G(s_2) \exp{\left(-s_2 r_{13}^2\right)}\,ds_2,	\\
	h(r_{23}) = \int_0^\infty H(s_3) \exp{\left(-s_3 r_{23}^2\right)}\,ds_3.
\end{gather}
\end{subequations}
Table \ref{tab:kern} contains kernels $F(s)$ for a variety of important two-electron operators $f(r_{12})$. \alert{From the formulae in Table \ref{tab:kern}, one can also easily deduce the Laplace kernels for related functions, such as $f(r_{12})^{2}$, $f(r_{12})/r_{12}$ and $\nabla^2f(r_{12})$}. Integrating over $\br_1$, $\br_2$ and $\br_3$ then yields
\begin{equation}
	[\bo] = S_1 S_2 S_3 \int_0^\infty \!\!\!\! \int_0^\infty \!\!\!\! \int_0^\infty F(s_1)\,  G(s_2)\, H(s_3)\, w_{\bo}(\bs) \,d\bs,
\end{equation}
where
\begin{equation} 
	w_{\bo}(\bs) =  \left[ \frac{\pi^3}{\zeta_1 \zeta_2 \zeta_3 D(\bs)} \right]^{3/2}  \exp \left[ -\frac{N(\bs)}{D(\bs)} \right],
\end{equation} 
and $\bs = (s_1,s_2,s_3)$.  The numerator and denominator are
\begin{subequations}
\begin{gather}
	N(\bs) 	= \kappa_{12} s_1 + \kappa_{13} s_2 + \kappa_{23} s_3
			+ \left(\frac{\kappa_{12}}{\zeta_3} + \frac{\kappa_{13}}{\zeta_2} + \frac{\kappa_{23}}{\zeta_1}\right) (s_1 s_2 + s_1 s_3 + s_2 s_3),			\\
	D(\bs)	= 1 + \left[\frac{1}{\zeta_1}+\frac{1}{\zeta_2}\right] s_1 + \left[\frac{1}{\zeta_1}+\frac{1}{\zeta_3}\right] s_2 + \left[\frac{1}{\zeta_2}+\frac{1}{\zeta_3}\right] s_3
			+ \frac{\zeta_1+\zeta_2+\zeta_3}{\zeta_1 \zeta_2 \zeta_3} (s_1 s_2 + s_1 s_3 + s_2 s_3),
\end{gather}
\end{subequations}
where
\begin{equation}
	\kappa_{ij} = \bY_{ij} \cdot \bY_{ij}
\end{equation}
is the squared length of the vector
\begin{equation}
	\bY_{ij} = \bZ_i - \bZ_j
\end{equation}
between the ith and jth gaussian product centers.

For reasons that will become clear later, it is convenient to introduce the generalized fundamental integral
\begin{equation}
	[\bo]^{\BM} = S_1 S_2 S_3 \int_0^\infty \!\!\!\! \int_0^\infty \!\!\!\! \int_0^\infty F(s_1)\, G(s_2)\, H(s_3)\, w_{\BM}(\bs) \,d\bs,
\end{equation}
where
\begin{equation}
	w_{\BM}(\bs) = \frac{s_1^{m_1} s_2^{m_2} s_3^{m_3}}{D(\bs)^{m_1+m_2+m_3}}
					\left[\frac{ s_1 s_2 + s_1 s_3 + s_2 s_3 }{D(\bs) }\right]^{m_4} w_{\bo}(\bs),
\end{equation}
and the auxiliary index vector $\BM=(m_1,m_2,m_3,m_4)$.

To form an $[a_1a_2a_3|b_1b_2b_3]$ class with a cyclic operator, we require all $[\bo]^{\BM}$ with
\begin{subequations} \label{eq:numm}
\begin{gather}
	0 \le m_1 \le a_1 + a_2 + b_1 + b_2,			\\
	0 \le m_2 \le a_1 + a_3 + b_1 + b_3,			\\
	0 \le m_3 \le a_2 + a_3 + b_2 + b_3,			\\
	0 \le m_4 \le a_1 + a_2 + a_3 + b_1 + b_2 + b_3.
\end{gather}
\end{subequations}
To form an $[a_1a_2a_3|b_1b_2b_3]$ class with a chain operator, the ranges of $m_1$, $m_2$ and $m_4$ are as in \eqref{eq:numm}, but $m_3 = 0$.

To construct an $\braket{aa|aa}$ class of two-electron integrals, one needs only $O(a)$ $[\bo]^{(m)}$ integrals.\cite{Review94}  However, it follows from \eqref{eq:numm} that, to construct an $\braket{aaa|aaa}$ class of three-electron integrals, we need $O(a^3)$ (for a chain operator) or $O(a^4)$ (for a cyclic operator) $[\bo]^{\BM}$ integrals.  This highlights the importance of computing these $[\bo]^{\BM}$ efficiently.  If at least one of the two-electron operators is a GTG, the $[\bo]^{\BM}$ can be found in closed-form. \cite{Boys60, Singer60}
Otherwise, they can be reduced to one- or two-dimensional integrals, which can then be evaluated by various numerical techniques.
This step can consume a significant fraction of the total computation time\cite{ssss91} and a comprehensive treatment of suitable numerical methods merits a detailed discussion which we will present in Part III of this series. \cite{3ERI3}

\subsection{Step T$_1$.  Build momentum on center $\bA_3$}
Given a set of $[\bo]^{\BM}$, integrals of higher angular momentum can be obtained recursively, following Obara and Saika. \cite{Obara86, Obara88}
Whereas VRRs for two-electron integrals have been widely studied, VRRs for three-electron integrals have not, except for GTGs. \cite{Persson97, Tenno00, Saito01,DahleThesis, Dahle2007,Dahle2008, Komornicki11} \phantom{x}

The T$_1$ step generates $[\ba_3]^{\BM}$ from $[\bo]^{\BM}$ via the 8-term VRR (see the Appendix for a detailed derivation)
\begin{align} \label{eq:T1_cyclic}
	[\ba_3^+]^{\BM}& = (\bZ_3-\bA_3) [\ba_3]^{\{0\}} + \zeta_1\zeta_2 \bY_{13} [\ba_3]^{\{2\}}
						+ \zeta_1\zeta_2 \bY_{23} [\ba_3]^{\{3\}} + (\zeta_1 \bY_{13} + \zeta_2 \bY_{23}) [\ba_3]^{\{4\}}		\notag	\\
					& + \frac{\ba_3}{2\zeta_3} \left\{	[\ba_3^-]^{\{0\}} - \zeta_1\zeta_2 [\ba_3^-]^{\{2\}}
						- \zeta_1\zeta_2 [\ba_3^-]^{\{3\}} - (\zeta_1+\zeta_2) [\ba_3^-]^{\{4\}} \right\},
\end{align}
where the superscript $+$ or $-$ denotes an increment or decrement of one unit of cartesian angular momentum.  (Thus, $\ba^\pm$ is analogous to $\ba\pm\bm{1}_i$ in the notation of Obara and Saika.)  The value in the curly superscript indicates which component of the auxiliary index vector $\BM$ is incremented.

For a chain operator, the $\{3\}$ terms disappear, yielding the 6-term RR
\begin{align} \label{eq:T1_chain}
	[\ba_3^+]^{\BM}& = (\bZ_3-\bA_3) [\ba_3]^{\{0\}} + \zeta_1\zeta_2 \bY_{13} [\ba_3]^{\{2\}} + (\zeta_1 \bY_{13} + \zeta_2 \bY_{23}) [\ba_3]^{\{4\}}	\notag	\\
					& + \frac{\ba_3}{2\zeta_3} \left\{	[\ba_3^-]^{\{0\}} - \zeta_1\zeta_2 [\ba_3^-]^{\{2\}} - (\zeta_1+\zeta_2) [\ba_3^-]^{\{4\}} \right\}.
\end{align}
It is satisfying to note that, by setting $\zeta_2 = 0$ in \eqref{eq:T1_chain}, we recover the Obara-Saika two-electron RR.

\subsection{Step T$_2$. Build momentum on center $\bA_2$}
The T$_2$ step forms $[\ba_2\ba_3]$ from $[\ba_3]^{\BM}$ via the 10-term RR
\begin{align} \label{eq:T2_cyclic}
  	[\ba_2^+\ba_3]^{\BM}	& = (\bZ_2-\bA_2) [\ba_2\ba_3]^{\{0\}} + \zeta_1\zeta_3 \bY_{12} [\ba_2\ba_3]^{\{1\}}
							 	- \zeta_1\zeta_3 \bY_{23} [\ba_2\ba_3]^{\{3\}} + (\zeta_1 \bY_{12} - \zeta_3 \bY_{23}) [\ba_2\ba_3]^{\{4\}}				\notag	\\
							& + \frac{\ba_2}{2\zeta_2} \left\{	[\ba_2^-\ba_3]^{\{0\}} - \zeta_1\zeta_3 [\ba_2^-\ba_3]^{\{1\}}
																- \zeta_1\zeta_3 [\ba_2^-\ba_3]^{\{3\}} - (\zeta_1+\zeta_3) [\ba_2^-\ba_3]^{\{4\}}	\right\}	\notag	\\
							& + \frac{\ba_3}{2} \left\{ \zeta_1 [\ba_2\ba_3^-]^{\{3\}} + [\ba_2\ba_3^-]^{\{4\}} \right\}.
\end{align}
For a chain operator, the $\{3\}$ terms disappear, yielding the 7-term VRR
\begin{align} \label{eq:T2_chain}
  	[\ba_2^+\ba_3]^{\BM}	& = (\bZ_2-\bA_2) [\ba_2\ba_3]^{\{0\}} + \zeta_1\zeta_3 \bY_{12} [\ba_2\ba_3]^{\{1\}} + (\zeta_1 \bY_{12} - \zeta_3 \bY_{23}) [\ba_2\ba_3]^{\{4\}}	\notag	\\
							& + \frac{\ba_2}{2\zeta_2} \left\{	[\ba_2^-\ba_3]^{\{0\}} - \zeta_1\zeta_3 [\ba_2^-\ba_3]^{\{1\}} - (\zeta_1+\zeta_3) [\ba_2^-\ba_3]^{\{4\}} \right\}
							    + \frac{\ba_3}{2} [\ba_2\ba_3^-]^{\{4\}}.
\end{align}

\subsection{Step T$_3$. Transfer momentum to center $\bA_1$}
The T$_3$ step generates $[\ba_1\ba_2\ba_3]$ from $[\ba_2\ba_3]$.  There are two possible ways to do this.  The first, which we call Step T$_3'$, is to build angular momentum directly on $\bA_1$ using the 12-term VRR (Eq.~\eqref{eq:T3_prime}).  However, this is computationally expensive and we choose to avoid it.  A second option exploits the translational invariance
\begin{equation}
	\sum_{j=1}^3 (\nabla_{\bA_j}+\nabla_{\bB_j}) [\ba_1\ba_2\ba_3] = \bo
\end{equation}
to derive the 6-term TRR
\begin{align} \label{eq:T3_cyclic}
	[\ba_1^+\ba_2\ba_3]	& = \frac{\ba_1}{2\zeta_1} [\ba_1^-\ba_2\ba_3] + \frac{\ba_2}{2\zeta_1} [\ba_1\ba_2^-\ba_3] + \frac{\ba_3}{2\zeta_1} [\ba_1\ba_2\ba_3^-]	\notag	\\
							& - \frac{\zeta_2}{\zeta_1} [\ba_1\ba_2^+\ba_3] - \frac{\zeta_3}{\zeta_1} [\ba_1\ba_2\ba_3^+]
								- \frac{\beta_1(\bA_1-\bB_1) + \beta_2(\bA_2-\bB_2) + \beta_3(\bA_3-\bB_3)}{\zeta_1} [\ba_1\ba_2\ba_3],
\end{align}
which transfers momentum between centers that host different electrons.

\subsection{Step C. Contraction}
At this stage, following the HGP algorithm, \cite{HGP88} we contract the $[\ba_1\ba_2\ba_3 | \bo\bo\bo]$ to form the $\braket{\ba_1\ba_2\ba_3 | \bo\bo\bo}$.
We can perform the contraction at this point because all of the subsequent RRs are independent of the contraction coefficients and exponents.
More details about this contraction step can be found in Ref.~\citenum{Review94}.

\subsection{Steps T$_4$ to T$_6$. Shift momentum to ket centers}
We shift momentum to $\bB_3$, $\bB_2$ and $\bB_1$ from $\bA_3$, $\bA_2$ and $\bA_1$, respectively, using the 2-term HRRs
\begin{align}
	\braket{\ba_1\ba_2\ba_3 | \bb_3^+}	& = \braket{\ba_1\ba_2\ba_3^+ | \bb_3} + (\bA_3-\bB_3) \braket{\ba_1\ba_2\ba_3 | \bb_3},										\label{eq:T4}	\\
	\braket{\ba_1\ba_2\ba_3 | \bb_2^+\bb_3}	& = \braket{\ba_1\ba_2^+\ba_3 | \bb_2\bb_3} + (\bA_2-\bB_2) \braket{\ba_1\ba_2\ba_3 | \bb_2\bb_3},					\label{eq:T5}	\\
	\braket{\ba_1\ba_2\ba_3 | \bb_1^+\bb_2\bb_3}	& = \braket{\ba_1^+\ba_2\ba_3 | \bb_1\bb_2\bb_3} + (\bA_1-\bB_1) \braket{\ba_1\ba_2\ba_3 | \bb_1\bb_2\bb_3}.	\label{eq:T6}
\end{align}

\section{Concluding remarks}
In this Article, we have presented a general algorithm to construct three-electron integrals over gaussian basis functions of arbitrary angular momentum from fundamental (momentumless) integrals. 
The algorithm is based on vertical, transfer and horizontal RRs in the spirit of the Head-Gordon-Pople algorithm.
In Part II of this series, we will report detailed investigations of upper bounds on these integrals.  In Part III, we will discuss efficient methods for computing the fundamental integrals. 
Our approach can be extended to four-electron integrals, and we will also report results on this soon.

\begin{acknowledgement}
P.M.W.G.~and P.F.L.~thank the National Computational Infrastructure (NCI) for supercomputer time. 
P.M.W.G.~thanks the Australian Research Council for funding (Grants No.~DP140104071 and DP160100246). 
P.F.L. thanks the Australian Research Council for a Discovery Early Career Researcher Award (Grant No.~DE130101441) and a Discovery Project grant (DP140104071).
\end{acknowledgement}
\vspace{10mm}

\appendix
\section{\label{app:RR} Derivation of VRRs}
In this Appendix, we follow the Ahlrichs approach\cite{Ahlrichs2006} to derive a VRR for the construction of $[\ba_1]^{\BM}$ integrals.

Defining the scaled gradient operator 
\begin{equation}
	\Hat{D}_{\bA_1}= \frac{\nabla_{\bA_1}}{2\alpha_1},
\end{equation}
we can write the Boys relation\cite{Boys50}
\begin{equation}
	[\ba_1^+] = \Hat{D}_{\bA_1} [\ba_1] + \frac{\ba_1}{2\alpha_1} [\ba_1^-],
\end{equation}
which connects an integral of higher momentum to an integral derivative with respect to a coordinate of $\bA_1$.

In operator form, this can be written as
\begin{equation} \label{eq:der3}
	[\ba_1^+] = \Hat{M}_{\ba_1} \Hat{D}_{\bA_1} [\bo] + \frac{\ba_1}{2\alpha_1} [\ba_1^-].
\end{equation}
Substituting the chain rule expression
\begin{equation}
	\Hat{D}_{\bA_1} = \frac{\nabla_{\bA_1}S_1}{2\alpha_1} \frac{\partial}{\partial S_1}
					+ \frac{\nabla_{\bA_1}\kappa_{12}}{2\alpha_1} \frac{\partial}{\partial \kappa_{12}}
					+ \frac{\nabla_{\bA_1}\kappa_{13}}{2\alpha_1} \frac{\partial}{\partial \kappa_{13}}
\end{equation}
into \eqref{eq:der3}, using the identities
\begin{gather}
	\frac{\partial [\bo]^{\BM}}{\partial S_1} = S_1^{-1} [\bo]^{\{0\}},														\\
	\frac{\partial [\bo]^{\BM}}{\partial \kappa_{12}} = - \zeta_1\zeta_2\zeta_3 [\bo]^{\{1\}} - \zeta_1\zeta_2 [\bo]^{\{4\}},	\\
	\frac{\partial [\bo]^{\BM}}{\partial \kappa_{13}} = - \zeta_1\zeta_2\zeta_3 [\bo]^{\{2\}} - \zeta_1\zeta_3 [\bo]^{\{4\}},
\end{gather}
the commutator property
\begin{equation}
	\Hat{M}_{\ba_1} p = p\,\Hat{M}_{\ba_1} + \ba_1\rho\,\Hat{M}_{\ba_1^-},
\end{equation}
(where $p$ is linear in $\bA_1$ and $\rho$ is its derivative) and the identities
\begin{gather}
	\Hat{D}_{\bA_1}(\bZ_1-\bA_1) = -\frac{\beta_1}{2\alpha_1\zeta_1},		\\
	\Hat{D}_{\bA_1} \bY_{12} = \Hat{D}_{\bA_1} \bY_{13} = \frac{1}{2\zeta_1}, 
\end{gather}
one eventually obtains the 8-term RR for building on $\bA_1$
\begin{align}
	[\ba_1^+]^{\BM}& = (\bZ_1-\bA_1) [\ba_1]^{\{0\}} - \zeta_2\zeta_3 \bY_{12}[\ba_1]^{\{1\}}
						- \zeta_2\zeta_3 \bY_{13}[\ba_1]^{\{2\}} - (\zeta_2 \bY_{12} + \zeta_3 \bY_{13}) [\ba_1]^{\{4\}}	\notag	\\
					& + \frac{\ba_1}{2\zeta_1} \left\{	[\ba_1^-]^{\{0\}} - \zeta_2\zeta_3[\ba_1^-]^{\{1\}}	
						- \zeta_2\zeta_3[\ba_1^-]^{\{2\}} - (\zeta_2 + \zeta_3) [\ba_1^-]^{\{4\}} \right\}.
\end{align}
Equation \eqref{eq:T1_cyclic}, which builds on $\bA_3$, can be derived similarly.  
However, for chain operators, the RR that builds on $\bA_1$ does not shed terms and it is therefore cheaper to build on $\bA_3$ than on $\bA_1$.

Equation \eqref{eq:T2_cyclic} can be derived using the relation
\begin{equation}
	[\ba_2^+\ba_3]^{\BM} = \Hat{M}_{\ba_3} [\ba_2^+]^{\BM},
\end{equation}
and the 12-term VRR for building on $\bA_1$ in Step T$_3'$ is
\begin{align} \label{eq:T3_prime}
	[\ba_1^+\ba_2\ba_3]^{\BM}	& = \Hat{M}_{\ba_3} \Hat{M}_{\ba_2} [\ba_1^+]^{\BM}														\notag \\
								& = (\bZ_1-\bA_1) [\ba_1\ba_2\ba_3]^{\{0\}}	
								- \zeta_2\zeta_3 \bY_{12} [\ba_1\ba_2\ba_3]^{\{1\}} 														\notag	\\
								&- \zeta_2\zeta_3 \bY_{13} [\ba_1\ba_2\ba_3]^{\{2\}}		
								 - (\zeta_2 \bY_{12} +\zeta_3 \bY_{13}) [\ba_1\ba_2\ba_3]^{\{4\}}											\notag	\\
								& + \frac{\ba_1}{2\zeta_1} \left\{	[\ba_1^-\ba_2\ba_3]^{\{0\}} - \zeta_2\zeta_3 [\ba_1^-\ba_2\ba_3]^{\{1\}}
									- \zeta_2\zeta_3 [\ba_1^-\ba_2\ba_3]^{\{2\}} - (\zeta_2 +\zeta_3)[\ba_1^-\ba_2\ba_3]^{\{4\}} \right\}		\notag	\\
								& + \frac{\ba_2}{2} \left\{ \zeta_3 [\ba_1\ba_2^-\ba_3]^{\{1\}} + [\ba_1\ba_2^-\ba_3]^{\{4\}} \right\}
									+ \frac{\ba_3}{2} \left\{ \zeta_2 [\ba_1\ba_2\ba_3^-]^{\{2\}} + [\ba_1\ba_2\ba_3^-]^{\{4\}} \right\}.
\end{align}
\bibliography{3eRR}

\end{document}